\documentclass[runningheads,envcountsame,envcountsect,oribibl]{llncs}

\usepackage[latin1]{inputenc}
\usepackage[colorlinks=true,ps2pdf=true]{hyperref}
\usepackage[dvips]{graphics}
\usepackage{epsfig}
\usepackage{psfrag}
\usepackage{amsmath}
\usepackage{amssymb}
\usepackage[english]{babel}
\usepackage{xspace}
\usepackage{url}
\usepackage{graphicx}

\makeatletter

\def\rcs$#1${#1}
\newcommand{\version}[1]{\thanks{\rcs#1}}

\newcommand{\id}[1]{\ensuremath{\mathit{#1}}}	% name of variable or identifier

\newcommand{\ie}{{i.e.},\ }		% i.e.
\newcommand{\eg}{{e.g.},\ }		% e.g.
\newcommand{\etal}{\textit{et al.}~}	% et al.

\newcommand{\seckey}[1]{{k_{\id{#1}}}}
\newcommand{\pubkey}[1]{{K_{\id{#1}}}}
\newcommand{\privkey}[1]{\seckey{#1}}
\newcommand{\encrypt}{\@ifstar{\@encryptshort}{\@encryptlong}}
\newcommand{\@encryptlong}[2]{E_{\seckey{#1}}(#2)}
\newcommand{\@encryptshort}[2]{\{#2\}_{\seckey{#1}}}
\newcommand{\pkencrypt}{\@ifstar{\@pkencryptshort}{\@pkencryptlong}}
\newcommand{\@pkencryptlong}[2]{E_{\pubkey{#1}}(#2)}
\newcommand{\@pkencryptshort}[2]{\{#2\}_{\pubkey{#1}}}
\newcommand{\sign}{\@ifstar{\@signshort}{\@signlong}}
\newcommand{\@signlong}[2]{S_{\privkey{#1}}(#2)}
\newcommand{\@signshort}[2]{[#2]_{\privkey{#1}}}
\newcommand{\mac}{\@ifstar{\@macshort}{\@maclong}}
\newcommand{\@maclong}[2]{\id{MAC}_{\seckey{#1}}(#2)}
\newcommand{\@macshort}[2]{[#2]_{\seckey{#1}}}
\makeatother

\newcommand{\SOD}{\id{SO}_D}
\newcommand{\SOLDS}{\id{SO}_{LDS}}

\title{Crossing Borders: Security and Privacy Issues of the European 
e-Passport\version{$Id: passport.tex,v 1.44 2006/06/30 07:25:14 ronny Exp $}
}

\author{Jaap-Henk Hoepman,
  Engelbert Hubbers,
  Bart Jacobs,
  Martijn Oostdijk,
  Ronny Wichers Schreur}
\institute{
Institute for Computing and Information Sciences \\
  Radboud University Nijmegen\\
  P.O. Box 9010, 6500 GL \ Nijmegen, 
  the Netherlands \\
  \email{\{jhh,hubbers,bart,martijno,ronny\}@cs.ru.nl}
}

\titlerunning{Crossing Borders}
\authorrunning{Hoepman \etal}

\begin{document}

\maketitle

\bibliographystyle{plain}

\begin{abstract}
The first generation of European e-passports will be issued in 2006.  We
discuss how borders are crossed regarding the security and privacy erosion of
the proposed schemes, and show which borders need to be crossed to improve the
security and the privacy protection of the next generation of e-passports.  In
particular we discuss attacks on Basic Access Control due to the low entropy of
the data from which the access keys are derived, we sketch the European
proposals for Extended Access Control and the weaknesses in that scheme, and
show how fundamentally different design decisions can make e-passports
more secure.
\end{abstract}

\section{Introduction} 

After several years of preparation, many countries start
issuing e-pass\-ports with an embedded chip holding biometric data of
the passport holder in 2006. This is a major ICT-operation, involving many
countries, most of them providing their own implementation, using
biometrics at an unprecedented scale. Passport security must
conform to international (public) standards, issued
by the International Civil Aviation Organization
(ICAO)~\cite{icao04:pki,icao04:lds}. The standards cover confidentiality,
integrity and authenticity of the passport data, including the facial
image. Additionally, the European Union (EU) has developed its own
standards (called ``Extended Access Control'').
% for the protection of
%fingerprints.

The present paper reviews these developments (like
in~\cite{juels2005passports,kc2005mrtd}) especially from a European
perspective, with corresponding emphasis on fingerprint protection.
Also it tries to put these developments within a wider perspective of
identity management (IM) by governments,
following~\cite{hoepman2006epassports}. This leads to a ``revision''
plan for e-passports.

From an academic background we, the authors, closely follow the
introduction of the e-passport in the Netherlands.
We have advised
the government on several matters, and are involved in public debates
on related issues. We have received an early test version of
the e-passport, and developed our own reader-side software, based on
the ICAO protocols. We have had access to confidential material
regarding the EU-protocols. However, the present paper is based solely
on publicly available material, and is organised as follows. 

We first discuss the main security
requirements the new e-passport should satisfy. After a brief discussion
of biometry in Sect.~\ref{sec-bio}, we
describe the standard security measures of the ICAO standard 
and the weaknesses associated with them in Sect.~\ref{sec-standard}.
Future European e-passports will be equipped with Extended Access Control, which
we outline in Sect.~\ref{sec-eac}, and whose shortcomings we also study.
e-Passports enable new applications. Sect.~\ref{sec-newapps}
discusses the danger of such function creep but also investigates 
the new possibilities created by such applications.
We study identity management issues of the e-passport in Sect.~\ref{sec-im},
and evaluate the realisation of the original goals in Sect.~\ref{sec-eval}.
We finish the paper with some proposals for more fundamental changes to
the architecture of a second generation of e-passports that will increase 
both their security and their flexibility of use in new applications.

\section{Aims and Security Goals}
\label{sec-aims}

It is a fact that modern passports are hard to forge. Thus, many
criminal organisations do not even try such fraud, but instead collect
large numbers of genuine passports, and pick one that shows a
reasonable resemblance to a member that needs a new
identity. Similarly, passports are sometimes borrowed for illegal
border crossing, and later returned to the rightful owner.

The original aim of the use of biometrics in travel documents is thus
to combat ``look-alike'' fraud. Hence the emphasis is on biometric
\textit{verification} (instead of \textit{identification}), involving
a 1:1 check to make sure that a particular passport really belongs to
a particular person.

The biometrics of the passport holder will be included in a chip that
is embedded in the passport. Communication with the chip will be
wireless, and not via contact points, because wireless communication
allows higher data rates, does not involve wear, and does not require
a change of the standard format of the passport to for instance
credit-card size\footnote{A change of format for other official
documents, like a drivers licence, is seen as less problematic, because such a
document is not stamped.}.
% Electronic stamps are not foreseen in the
%e-passport, because the chip is sealed after issuance: no information
%can be added or changed for security reasons.}.
%
% CHANGED: actually, VISA's are foreseen to be added to the datapage 
% of the passport

The wireless character does introduce new security risks
(with respect to traditional passports), for the holder, the issuing state, and for
the accepting state. At a high level of abstraction, the following
three security goals seem reasonable. The first two focus on 
confidentiality for the passport holder. The last one mainly
concerns the accepting (and also issuing) state.
\begin{enumerate}
\item A passport reader should identify 
itself first, so that only ``trusted'' parties get to read the
information stored in the chip.

\item No identifying information should be released without consent
of the passport holder.

\item The receiver of the information should be able to establish
the integrity and authenticity of the data.
\end{enumerate}

The first goal relates to the situation where for instance a police
officer wishes to check your identity. In most countries you have the
right to ask the police officer in question to identify himself first,
so that you can be sure that you are dealing with a genuine
representative of the state. The second goal is relevant to prevent
``RFID-bombs''~\cite{juels2005passports} for instance, that are
activated by the immediate presence of (the passport of) a particular
person, or citizen of a particular country. 
%A more subtle issue is that even differences between countries
%should not be recognisable. 
Such information is also useful for a
terrorist who is trying to decide whether to blow himself up in a
particular bus.  
%The third security goal speaks for itself.
%
We shall evaluate the realisation of these goals later on,
in Sect.~\ref{sec-eval}

\section{Biometry}
\label{sec-bio}

This paper does not focus on the biometry involved, but a few words
are in order. ICAO has opted for the use of facial images and
fingerprints as primary biometrics because they are reasonably
familiar, easy to use, and non-intrusive. A controversial issue---from
a privacy perspective---is that the passport chip will not contain
templates but pictures (actual JPEGs). The reason is that there is no
well-established digital standard for such templates, and early
commitment to a closed proprietary format is not desirable. This means
that if a passport chip (or data base, or reader) is compromised,
original biometric data leaks out, which may lead to reconstruction
and additional (identity) fraud.

The effectiveness of biometry is highly overrated, especially by
politicians and policy makers. Despite rapid growth in applications,
the large-scale use of biometry is untested. The difficulty is that it
is not only unproven in a huge single application (such as
e-passports), but also not with many different applications in
parallel (including ``biometry for fun''). The interference caused by
the diversity of applications---each with its own security policy, if
any---may lead to unforeseen forms of fraud.

A basic issue that is often overlooked is fallback. What if my
biometric identity has been compromised, and I am held responsible for
something I really did not do, how can I still prove ``it wasn't me''?

The Netherlands has recently conducted a field test for the enrolment
procedures of the biometric passport, see~\cite{ToBeOrNotToBe},
involving almost 15.000 participants.  The precise interpretation of
the outcome is unclear, but failure-to-acquire turns out to be a
significant problem, especially for young and elderly
people. Substantial numbers of people will thus not have appropriate
biometric travel documents, so that fully automatic border crossing is
not an option.

\section{Standard Security Measures (ICAO)}
\label{sec-standard}

The various ICAO standards for machine readable travel documents,
notably \cite{icao04:pki} and \cite{icao04:lds}, specify
precise requirements for accessing and interpreting the contents of
the embedded chip. Different security controls are described to ensure that
different security goals are met. We discuss these in the order in which the
mechanisms are used in a typical session between reader
(or: inspection system, the computer that is attempting to read information
from the document) and the European passport chip.

\begin{figure}[t]
\begin{center}
\includegraphics[scale=0.4]{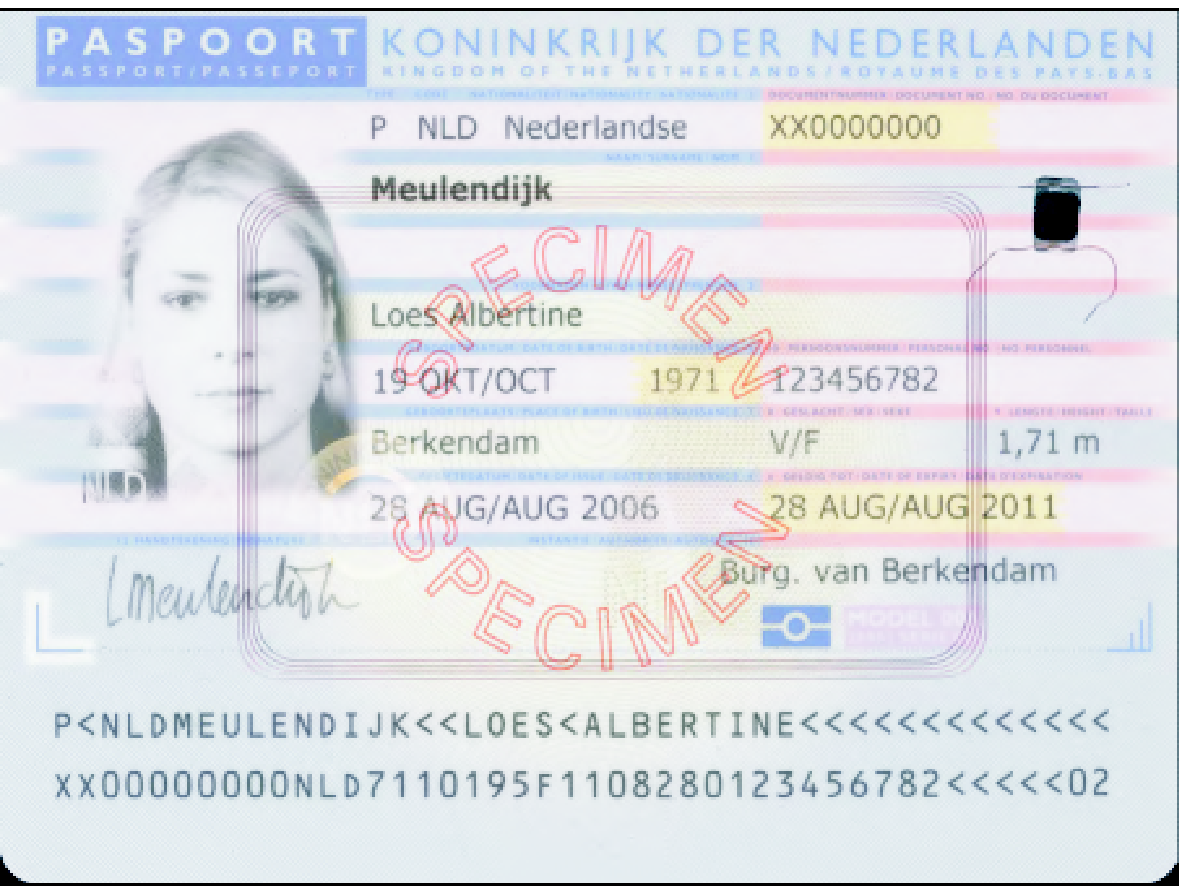}
\caption{Example of a Dutch passport. The two bottom lines of text are the
 MRZ.}
\label{fig-example}
\end{center}
\end{figure}

\paragraph{BAC: Basic Access Control.}
%% Optional in ICAO, mandatory in Dutch...
Before any information can be read from a passport, the reader needs to
go through \emph{basic access control\/} (BAC). This is a challenge-response
protocol in which the reader proves to the passport that it has knowledge of
the contents of the machine readable zone (MRZ). The MRZ consists of two lines
of optically readable text containing among others the name of the holder,
and the passport number. It is printed on the first page of
the physical document (See Fig.~\ref{fig-example}).

The procedure is as follows. The reader optically reads the contents of the
MRZ, and derives the \emph{access key} seed
$\seckey{IFD/ICC}$ from the data it reads.
After that, the reader proves to the chip that it has optically read the MRZ by
signing a random challenge from the chip using a key derived from the access
key seed.
Subsequently, 
passport and inspection system exchange some extra random data, which is then
used to generate session keys and an initial counter for secure messaging.
The session keys are fresh for each session.

BAC prevents so-called \emph{skimming\/} of passports, \ie reading the
contents without the cardholder's knowledge. Note that BAC does not
authenticate the reader: anyone who knows the MRZ can successfully
complete BAC and continue reading other information on the chip.

\paragraph{SM: Secure Messaging.}

%% Mandatory if BAC is present?
Confidentiality and integrity of all communication between reader and passport
is provided by so-called \emph{secure messaging}.
Commands sent to the passport as well as responses
sent back to the reader are encrypted and augmented with a message
authentication code (MAC), using the keys established during BAC. 
A sequence counter is included to prevent replay of messages.

\paragraph{PA: Passive Authentication.}

%% The inspection system can now attempt to read the contents of the
%% passport's memory.
The data stored on the passport is organised in a logical data structure (LDS),
which consists of a number of files (called data groups). Typical examples of data
groups are: a file containing the information in the MRZ, a file containing a
JPEG image of the cardholder's face, and files containing other biometric
features such as the cardholder's fingerprints.

%% SOd: Document Security Object
Each data group in the LDS is hashed. All these hashes together form
the (document) \emph{security object} $\SOLDS$. The security object
is signed by the issuing country and the result, $\SOD$, is stored on
the passport as well. This means that
the inspection system can check that the contents of the LDS have not been
altered during communication, thus ensuring the integrity of the LDS. The
standards refer to this integrity protection mechanism as
\emph{passive authentication}.

\paragraph{AA: Active Authentication.}

%% Optional in ICAO, will be in Dutch.
To prevent cloning of the chip, an integrity mechanism called
\emph{active authentication} is used, in which the passport
proves possession of a private key $\privkey{AA}$ using a
challenge-response
protocol. The corresponding public key, needed by the inspection
system to check the response of the passport, is part of the LDS
and can be read by the inspection system. A hash of this public
key is signed through the $\SOD$, to ensure authenticity. 
%Active authentication makes it
%impossible to create a copy of a passport, unless of course the
%private key is compromised somehow.

\subsection{Guessing the Access Key}
\label{subsec-guess}

To access the passport without having its MRZ, one needs to guess the access
key seed $\seckey{IFD/ICC}$, which is $128$ bits long.  The National Institute
of Standards and Technology (NIST)~\cite{nist-keys} and the ECRYPT EU Network
of Excellence on cryptology~\cite{ecrypt-keylength} recommend $80$ bits for a
minimal level of general purpose protection in 2005, and $112$ bits ten years
from now.  In other words, the access key seed is long enough to provide
adequate security.

But the fact that the access key seed is derived from information in the MRZ can be
used to the attackers' advantage.  The `MRZ-information'
consists of the concatenation of the passport number, date of birth
and date of expiry, including their respective check digits, as described
in~\cite{icao03:doc9303}. Given a guess for the MRZ-information, the
corresponding access key seed $\seckey{IFD/ICC}$ is easily calculated, and from
that all other session keys can be derived as well. These keys can then be
tried against a transcript of an eavesdropped communication session between
this passport and the reader, to see if they deliver meaningful data. 
%If so,
%the guess of the MRZ-information was correct, and the whole session is
%compromised. 

To estimate the amount of work the attacker needs to perform for
such an off-line attack, we estimate
the amount of Shannon entropy of each of these fields. We should stress this is
a very crude approach (unless we assume the underlying probability
distributions are uniform). For lower bounds, we should in fact use the
Guessing entropy~\cite{massey1994guessingentropy} ($\sum_i i p_i$) or even the
min-entropy ($\min_i - \log p_i$). The Shannon
entropy only gives us an upper bound, but if that bound is small
the security of the system is most certainly weak.

The entropy of the
\emph{date of birth} field is $\log ( 100 \times 365.25) = 15.16$ bits,
as it can contain only the last two digits of the year of birth.
If one can see the holder of the passport and guess his age correct within a
margin of $5$ years, the entropy of this field decreases to $10.83$.

The \emph{date of expiry} is determined by the date of issuing and the validity
period of a passport. In the Netherlands, passports are valid for $5$ years,
and are issued only on working days (barring exceptional circumstances).
For a \emph{valid} passport, the entropy of this field becomes
$\log ( 5 \times 365.25 \times 5/7 ) = 10.34$.

The MRZ field for the \emph{passport number} can contain $9$ characters. If the
passport number is longer, the excess characters are stored in the MRZ optional
data field (which is not used to derive the access key seed).
The entropy of the passport number
field, assuming digits and upper-case letters only, becomes
$\log((26+10)^9)=46.53$.  Many countries have further restrictions on the
format of their passport numbers. Passport numbers may contain check digits, or
start with a common prefix to distinguish passport types (\eg military
passports).

At best, the total entropy of date of birth, date of expiry and passport number
becomes $15.16 + 10.34 +46.53 = 72.03$, which is less than $80$ bits
recommended by both NIST and ECRYPT~\cite{ecrypt-keylength,nist-keys}
to protect against eavesdropping and other off-line attacks.
It \emph{is} sufficient to protect against
skimming attacks (where possible keys are tried on-line) because the passport 
is slow to respond to each individual key tried.

\begin{figure}[t]
\begin{center}
\includegraphics[width=10cm]{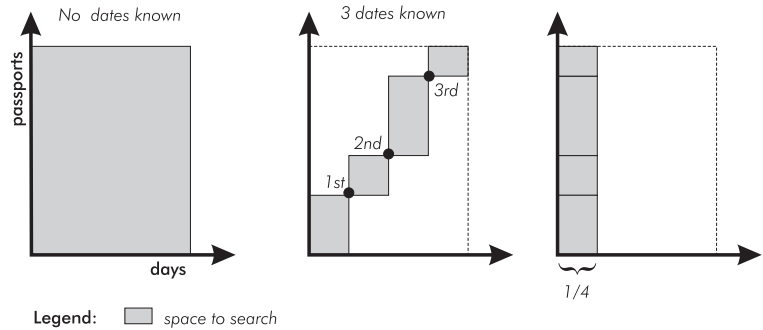}
\end{center}
\caption{Known dates of issuing reduce the search space.}
\label{fig-entropy}
\end{figure}

In certain countries the situation is even worse. Often, passport
numbers are issued sequentially. This implies there is a correlation between
the date of issue (and therefore date of expiry) and the passport number. Moreover,
all currently valid passports numbers (ignoring stolen or otherwise invalidated
ones) form a consecutive range, which is no longer than the total number of
people of that nationality. For the Dutch passport for instance,
bounding the population from above by $20$ million, the passport number entropy
drops to $\log(20 \times 10^6) = 24.25$.

With sequentially issued passports, the entropy drops even further with every
known combination of a passport number and the expiry date.
Suppose we know $k$ such combinations. This gives rise to
$k+1$ intervals of possible passport numbers for a given date range
Let us take the rather pessimistic approach that 
 we do not assume anything about the distribution of passports over
dates within those intervals (although it is very likely that passports are
issued at a reasonably constant rate).
On the optimistic side, let us assume the $k$ known passports are
issued evenly distributed over the validity period length.
This reduces the search space by a factor $k+1$ 
as illustrated in Fig.~\ref{fig-entropy}.
Hence the entropy 
of expiry date plus passport number drops with
$\log (k+1)$. 
For the Dutch passport, using $k=15$ and the figures above, the entropy 
of the passport number becomes as small as $20.25$, and the total
entropy could be as small as 
$10.83 + 10.34 + 20.25 = 41.42$
(when we assume we can guess the age of the passport holder).

One obvious idea is to include the MRZ optional
data field in the list of MRZ items that is used to derive the MRZ access
key seed. This would increase the entropy of the MRZ access key seed,
especially if this 
optional data field is filled with random data.
Unfortunately some countries 
already use this field for other purposes. In the Netherlands, for instance,
this field stores the social-fiscal number, which is uniquely linked to an
individual and not very secret information.
In fact, this idea was recently rejected for inclusion in the ICAO standards.

\subsection{Traceability} %Subliminal channels
\label{subsec-trace}

To avoid collisions, contactless smart cards and RFID systems 
use unique low-level tag identifiers in the radio communication
protocol. This is also true for the e-passports.
If this identifier is fixed (which is usually the case in RFID tags and
contactless smart cards), passports are clearly easily traceable.
Note that because this identifier is used in the very first stages of setting
up a connection between the passport and the reader, no form of access control
or reader authentication can be performed. 
%This identifier, by necessity, is
%broadcast to any reader that tries to set up a connection with a passport.

Luckily, this anti-collision identifier does not have to be fixed. The
number can also be randomly generated each time the passport comes within range
of a reader. 
If the random generator is of sufficient quality (and this is certainly an
issue in low-end RFID systems), the passport can no longer be traced
through the anti-collision identifier.

However, the anti-collision identifier creates a possible subliminal
channel. For instance, instead of simply generating a random number $r$,
the passport could be instructed to generate an anti collision identifier 
like
\[
   \id{id} = \encrypt{NSA}{r,\id{passportnumber}}~.
\]
The resulting string looks random, because of the randomness of $r$ and the
properties of the encryption function. But clearly it can be decrypted
by the owner of $\privkey{NSA}$ to reveal the passport number.
Unless the passport chip is reverse engineered, the existence of such a
subliminal channel cannot be detected.

Another subliminal channel exists when Active
Authentication is used~\cite{bsi2006extendedaccesscontrol}. 
Recall that active authentication requires the
passport to sign a challenge from a reader
using its unique private key.
Because the challenge is totally determined by the reader, the reader
can embed information into this string, which is unknowingly signed by the
passport. For instance, the challenge could contain the border crossing
location, and the current
date and time. A signature adds an extra layer
of non-repudiability to the border crossing logs, and can be used to prove this
fact to others. 
The challenge could also contain
the passport number of the person verified ahead of you at border inspection,
possibly linking you to the person you were travelling with.

Even if all the above issues are addressed,
discriminating features of passports remain. Different countries may
use different chip suppliers. Later batches of passports will use more
advanced technology, or may contain different or additional
information\footnote{Indeed, the first passports will be issued
without fingerprints.}.  In the future, newer versions
of chip operating systems may be used.  All these differences may be
noticeable by looking carefully at the behaviour of the chip on the
radio channel, at the chip's Answer To Reset (ATR), which is sent in
reaction to a reset command by the reader, or at the responses the
chip gives (or doesn't give) to specific card commands sent to it. We
expect to see large differences in behaviour especially on unintended,
unexpected or even unspecified input sent to the
card.
All these things are
possible before BAC has been performed.

Other applications may be put on the passports (see 
Sect.~\ref{sec-newapps}) as well.
These applications may even be accessible before BAC has been performed.
The set of available applications may actually constitute a narrow profile that
identifies a specific set of possible passport holders,
and may reveal the place of work, or the banks the passport
holder has accounts with.

We conclude that even without access to the MRZ, \ie in the classic skimming
scenario on streets, public transport, etc.,
passports still leak information that can be traced back to individuals,
or groups of individuals.

\section{Extended Access Control}
\label{sec-eac}

Standardisation of the security features and biometrics to be used in European
passports has been taken up independently (but of course in accordance with the
ICAO standards) by the European Union~\cite{COM(2004)0116}.
In recognition of the fact that biometric information is quite sensitive, the 
European Union has mandated that such data should be protected by a so-called
``Extended Access Control'' mechanism.
The technical specifications of the European e-passport are drafted by a
special EU Committee, founded as a result of
Article 6 of Regulation 1683/95 laying down a uniform
format for visas~\cite{COM(2001)577}.

Public information about the details of Extended Access Control 
has recently become
available~\cite{bsi2006extendedaccesscontrol,kuegler2005eac}. 
This allows us to discuss
certain shortcomings in the schemes under consideration, although we wish to
stress that these schemes are a huge improvement over the extremely minimal
security features imposed by the ICAO standards.

Extended Access Control consists of two phases, Chip Authentication followed
by Terminal Authentication. 
Chip Authentication performs the same function as
Active Authentication in the ICAO standards, \ie proving the chip is genuine
and thus protecting the passport against cloning. It avoids the problems
associated with active authentication, like the challenge semantics discussed
in the previous section. Chip authentication achieves its task by 
first exchanging a session key using a Diffie-Hellman key exchange.
The chip uses a static key pair for this, the public part of which is part of the
logical data structure (LDS) on the chip
and thus signed through the security objects $\SOD$. The terminal uses a fresh
key pair for each session. Authenticity of the chip is established once
the chip proves that it knows the session key, which happens implicitly
when the session key is used successfully to communicate with the chip.

\begin{figure}[t]
\begin{center}
\includegraphics[width=10cm]{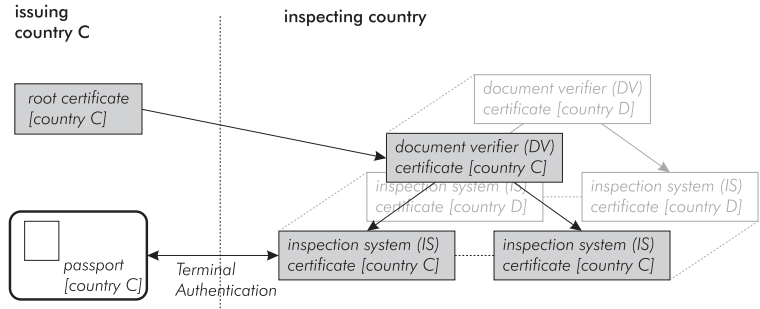}
\end{center} 
\caption{Extended Access Control certificates}
\label{fig-certs}
\end{figure}

Terminal Authentication aims to prove to the chip that the terminal is allowed
to access the data on the chip. This access is granted through a chain of
certificates, the root of which is the issuer of the passport at hand
(see Fig.~\ref{fig-certs}). In other words, the issuer of the passport
controls who can access the data on the passport.
This root issues Document Verifier (DV) Certificates, one for each country that is
granted access to the data on the passport. These DV certificates
are used to generate Inspection System (IS) certificates, which can be
distributed to inspection systems (\eg readers/terminals) at border crossings. 
Each passport issued by a particular country can verify the authenticity of
these DV certificates, and hence of the IS certificates issued through
these DV certificates. A valid IS certificate grants access to certain data on
the chip. All certificates have a limited validity period.

Terminal authentication, as proposed, does have a few weaknesses. 
First of all, the chip cannot keep time itself, and does not have access
to a reliable source of time either. This makes it hard to check whether a
certificate has expired or not. This, in turn, makes it practically impossible
to revoke a certificate. The problem is the following. A terminal with a valid
IS certificate and a valid DV certificate can access the sensitive data on 
many passports. When such a terminal is stolen, these access rights remain,
even when the validity period of these certificates has expired: the chip does
not know the correct time, and the terminal does not have to tell it the
correct time. This is the case even if certificates have extremely short
validity periods, like a single day. 
We see that one stolen terminal breaks the intended security goal of 
terminal authentication. Of course, stolen terminals do not make skimming
attacks possible: a terminal still needs access to the MRZ in order to 
perform basic access control. To mitigate the problem somewhat, the standards
propose that the chip keeps the most recent date seen on a valid
certificate. In other words, the chip advances its idea of the current time
each time it passes a border inspection system. This only saves the frequent
travellers; people that barely use their passports stay vulnerable for a long
time. 

Secondly, the certificate hierarchy itself poses a problem.
The hierarchy is quite shallow.
It does not
make it easy to allow access to the biometric data for other
applications beyond border inspection, even though such applications are
already being discussed today (see also Sect.~\ref{sec-newapps} below).
To acquire access, one has to apply for IS certificates at the country
DV, or for a DV certificate at each issuing country. The latter would create a
huge management overhead, as it would require each country to reliably verify
the identity and trustworthiness of the requesting applicant and issue
certificates in response. The first makes it impossible for countries to
differentiate access rights among different applications, and would make the
country DV responsible for the issuing of IS certificates for each and every
terminal involved in the new application. This is clearly impractical, if we
consider the use of passports for home banking or single sign-on systems that
require terminals at each and every PC.

Making the certificate hierarchy larger and more flexible may not be an
option. It means the
chip has to verify even more certificates before it can grant access. This
does put quite a burden on the processing capabilities of the chip,
which should guarantee reasonably short transaction times. No one is willing to
stand in the queue at border inspection for an even longer amount of time,
simply because the new passports contain new, but slow, technology.
A different, more flexible, approach is discussed in Sect.~\ref{sec-v2}.

\section{New Applications}
\label{sec-newapps}

The new e-passport requires an international infrastructure for
biometric verification. This is a huge project, of which the
effectiveness and risks are uncertain. The main driving force is
political pressure: the logic of politics simply requires high
profile action in the face of international terrorism.
Once implemented, it inevitably leads to
function creep: new possible applications emerge, either spontaneously
or via new policy initiatives. We shall discuss two such
applications, and speculate about the future.

%\subsection{E-passports for private use}

Once we obtain a new passport with a high-end chip embedded with
which we can communicate ourselves (via open standards), we can
ask ourselves whether we can also use it for our own purposes.
We briefly discuss two options: logon, and digital signatures.

The e-passport can be used to log on to your computer account.  For
instance, if you give your MRZ (or the associated keys) to your
computer or local network, the logon procedure can set up a
challenge-response session with your passport: activation of the chip
happens via the MRZ, and checking of a signature written by the
passport-chip on a challenge generated by your computer can proceed
via the public key of the document signer. It allows your computer to
check the integrity of the passports security object, which contains
the public key corresponding to the private signature-key of your
passport.

This authentication procedure only involves ``something you have'':
anyone holding your passport can log in to your machine. You can
strengthen the procedure by requiring a usual password, or even a
biometric check based on a comparison of facial images (a freshly
taken one, and the one on the chip).

You may also wish to use your e-passport to sign documents and
emails using the embedded private key for Active Authentication. 
This is not such a good idea, for two reasons.
First of all, the signature is obtained by exploiting the challenge-response
mechanism for another purpose. Such interference should be avoided,
because a challenge-response at a border inspection could then be
misused to trick you into signing a certain document.
Secondly, proving your identity after signing requires publication of your MRZ,
together with the security object of your passport-chip (which is
integrity-protected by a signature of the document signer): this
object couples the public key (for your signature) to identifying
information such as your name. But releasing the MRZ allows everyone
to access your passport, through Basic Access Control.

The underlying problem is that the e-passport was not designed with an
embedded useful certificate (such as X.509) for the holder.

%\subsection{Biometric identification}

Once there is an infrastructure for biometric verification, it becomes
natural to ask: why not use it for identification as well? People may
loose (willingly or unwillingly) their passport, or may apply for
multiple copies, possibly under different names. Indeed, the
government of the Netherlands is preparing
legislation~\cite{TK25764-26,TK29754-5} to set up a
central database with biometric information, in order to ``increase
the effectiveness of national identification laws''. Such a central
database goes beyond what is required by European directives.

The possibility of biometric identification of the entire
(passport-holding) population involves a change of power balance
between states and their citizens. Consent or cooperation is then no
longer needed for identification. Tracing and tracking of
individuals becomes possible on a scale that we have not seen before.

%\subsection{Speculation}

Assuming the biometric passport leads to a reliable infrastructure for
verification and identification of individuals, the societal pressure
will certainly increase to use it in various other sectors than just
border inspection. Such applications are not foreseen---or covered
by---European regulations. Interested parties are police and
intelligence forces, banks and credit card companies, social security
organisations, car rental firms, casinos, etc. Where do we
draw the line, if any?

We see that the introduction of the new e-passport is not only a large
technical and organisational challenge, but also a societal
one. Governments are implicitly asking for acceptance of this
new technology. This acceptance question is not so explicit, but
is certainly there. If some political action group makes a strong
public case against the e-passport, and manages to convince a large
part of the population to immediately destroy the embedded chip
after issuance---for instance by putting the passport in
a microwave---the whole enterprise will fail. The interesting
point is that individuals do have decisive power over
the use of the chip in their e-passport.
Even stronger, such a political action group may decide to
build disruptive equipment that can destroy the RFID-chips
from some distance, so that passports are destroyed without
the holder knowing (immediately).
%
%[Footnote: this should be done with some care, \eg in order not
%to kill all pacemakers withing range.]
To counter such movements, governments may try to make it
sufficiently unattractive or even impossible
to cross borders for travellers without a functioning passport. This
is only possible, however, if the numbers of broken chips is relatively low. 
And in any case, it will not improve popularity of the scheme to begin with.
%
%Also,
%if its population does not accept the e-passort it may decide to
%introduce more draconian measures, like setting-up a central database
%with biometric data.

\section{Identity Management Issues}
\label{sec-im}

Identity management (IM) is about ``rules-4-roles'': regulation of
identification, authentication and authorisation in and between
organisations. The new e-passport is part of IM by states. It forms an
identification and authentication mechanism that is forced upon
citizens, primarily for international movement, but also for internal
purposes. 

Identification and authentication in everyday life is a negotiation
process. When a stranger in the street asks for your biometric data, you
will refuse. But you may engage in a conversation, discover mutual
interests, and exchange business cards or phone numbers. Upon a next
contact more identifying information may be released, possibly leading
to a gradual buildup of trust.

The e-passport, in contrast, provides a rigid format. In certain
situations it forms an overkill, for instance when you just need to
prove that you are over eighteen. When IM goes digital and becomes
formalised one would like to have more flexible mechanisms, with
individual control via personal policies. In the future we may
expect to be carrying identity tokens that flexibly react to the
environment. Three basic rules for such systems are:
\begin{itemize}
\item The environment should authenticate itself first. For instance,
when the environment can prove to be my home, my policy allows my
token to release much personal information, for instance about my
music preference or health.

\item Authentication should be possible in small portions, for instance
via certificates or credentials saying ``this person is over
eighteen'', with a signature provided by a relevant authority.

\item Automatic recognition of individuals, for instance
via an implanted RFID chip that broadcasts your personal (social
security) number, is excluded.  
%This introduces too many risks, for
%instance via surgical personalised RFID-bombs. 
Privacy is important
for personal security---and not, as too often stated, only an
impediment to public security.
\end{itemize}

\section{Evaluation of Security Goals}
\label{sec-eval}

In Sect.~\ref{sec-aims} we have formulated three security goals that we
consider reasonable. In this section we evaluate whether the current system
meets these goals.

\paragraph{Readers should identify themselves first}

In the usual sense of ``authenticated'' or ``trusted'' readers, this goal
is not reached.
For instance,  we managed to write our own terminal application that retrieves 
the public information like the facial image from the chip. And our reader is 
not considered trusted.
The implemented BAC protocol only assures that the reader has knowledge of
the MRZ on the passport.
In the European implementation of EAC the reader must authenticate itself and
hence this goal is more or less met
for the information marked as sensitive,
but weaknesses exist (see Sect.~\ref{sec-eac}).

\paragraph{Consent by the passport holders}

Theoretically this goal is reached.
By use of BAC any terminal that tries to read information first needs to
read the MRZ information printed on the inside of the passport.
Hence the holder must give his consent for the transaction by opening his
passport.
However, as we have seen in Sect.~\ref{subsec-trace} some subliminal channels 
exist that may leak information about the card even before BAC has been
applied or in other words even before the holder has given his consent.

\paragraph{Proof of integrity and authenticity}
The integrity part of this goal is 
reached by the secure messaging system, which is applied for all
communication after BAC. As we have seen in Sect.~\ref{sec-standard} both
commands and responses are encrypted and augmented with a message
authentication code to provide integrity and confidentiality.
Authenticity of the information is guaranteed through Passive Authentication
(see Sect.~\ref{sec-standard})

\section{e-Passport v2}
\label{sec-v2}	

Until now we have discussed several issues with the security and privacy
protection of the current proposed standards for biometric passports, from both
ICAO and, in particular, the EU. We have argued that protection mechanisms
should be improved. However, improvements to such standards are at best
incremental, and do not usually challenge the primary design decisions. In
fact, such fundamental changes would certainly be backwards incompatible, and
require a totally new standard. In our opinion, more fundamental changes are
required to really provide strong security and proper privacy protection
to the new generation of e-passports.

\subsection{Avoiding Contactless Cards}

The most fundamental change is to reconsider the choice for a wireless
communication interface between the chip in the passport and the terminal at
border inspection. Using a wireless interface makes skimming attacks
possible. It is exactly the fear of this possibility that has sparked a huge
controversy over the current e-passport proposals. Initially, the US passports
would not even implement Basic Access Control. Now they are even considering to
include metal shields in the cover pages of the passport to function as a
Faraday cage, to physically disable the wireless communication link.

But all Basic Access Control really is, is a very elaborate way to achieve
exactly the same as what is achieved when inserting a smart card with contacts
into the slot of a reader: namely that the holder of the passport allows the
owner of the terminal to read the data on the chip. Then, why not simply use
smart cards with contacts for the new e-passport? The main arguments
against this have been the form-factor of the passport, and the need
for a sufficient bandwidth to quickly transmit the biometric data from the card
to the terminal. However, identity cards and drivers
licenses with dimensions similar to credit cards (ID-1) are already under
consideration. And bandwidth concerns
are no longer an issue either. 
Many smart card suppliers already sell smart cards with integrated USB 1.1
interfaces that allow for a much higher throughput, using the original~\cite{ISO7816} 
ISO contact module found on the card, and standardisation for this approach
is underway~\cite{ISO7816-12}.
Such a solution would take away all worries associated with using a
wireless chip, and would keep the e-passport clear of all discussions 
surrounding the (perceived) privacy issues with RFID.

\subsection{On-line Terminal Authentication}

Once a connection between passport and terminal is established, a decision has
to be made regarding the access rights of the terminal and to determine which
data on the passport it is allowed to read. Current EU proposals for 
extended access control
are found wanting: stolen terminals cannot be revoked, and the shallow, rigid
certificate hierarchy proposed to regulate access does not allow for flexible
and/or dynamic access control policies (see Sect.~\ref{sec-eac}).
The EU approach was chosen to allow
for off-line, mobile terminals, like those that are used by mobile
border inspection units. But clearly such mobile terminals can be connected to
the network over a wireless link, if only through GPRS, which is the standard
on cell phones these days.

If we assume that terminals are always connected to the network, we can use
on-line terminal authentication. The general idea is then the following.

Each terminal owns a private/public key pair. Each terminal is used for a
particular application. This application is encoded in a certificate $C_{AA}$
that contains the public key $K_{TA}$ of the terminal, and which is signed by
the application authority $AA$. Access rights are associated with
application. Each country stores, for each application authority that it wishes
to recognise, the access rights for that application. These access rights are
stored in the back office. The back office also stores the public keys of all
terminals that have been revoked.

On-line terminal authentication then proceeds as follows.  First, the terminal
sends the certificate $C_{AA}$ (containing its public key $K_{TA}$) to the
chip.  The 
chip and the terminal perform a challenge-response protocol in which the
terminal proves to the chip that it owns the private key corresponding to
$K_{TA}$. This establishes the identity of the terminal.  Next, the chip sets
up an authenticated channel between itself and the back office of the issuing
country. It can do so using a country certificate that is stored in the chip
during personalisation. The channel should not be vulnerable to replay attacks.
It sends $C_{AA}$ (and $K_{TA}$) 
to the back-office. There, $C_{AA}$ is verified against the known application
authorities (this validates that $K_{TA}$ was certified by such
an authority) and
$K_{TA}$ is checked against the list of all revoked
terminals. If these checks pass, the access rights for $AA$ are sent back to
the chip. If not, then the empty set (\ie no access rights) is sent back to
the chip.
The chip interprets the access rights it receives and grants access to the
terminal accordingly. Because the channel is authentic and does not allow
replay attacks, the access rights received by the chip correspond to the
certificate it sent to the back office.

With on-line terminal authentication, terminals can be revoked in real-time: as
soon as they are marked as revoked in the back offices of the issuing country,
no passport of that country will allow that terminal access to its data.
Also, the access permissions can be changed dynamically, and can even be based
on the exact time the request was made, or on the specific usage pattern of the
passport.
The general idea can be refined to also allow revocation of terminals by the
countries that manage them, instead of requiring them to inform all other
countries that a particular terminal should be revoked (because it was stolen,
for instance).
Also, more levels of certificates can be introduced, to make management of 
access rights easier.

\subsection{Other Improvements}

In Sect.~\ref{sec-bio} we have seen that real pictures are stored on the chip.
With an immediate consequence that whoever is able to retrieve these 
images from the chip, has access to good biometric data, which he can use
for identity theft. Using templates that work like a one-way function, it 
will be possible to check whether the template on the chip matches the
template derived from the person who is claiming to be the holder of the
passport.
This leaking of real biometric data may not seem such a big deal in a 
time where many pictures are published on the Internet.
The point here is that these pictures for the passports are taken under
good conditions and hence provides highly accurate biometric information.

The entropy-related off-line
attacks discussed in Sect.~\ref{subsec-guess} are possible
because a guess of MRZ-information directly leads to all keys used in a
communication session. These keys can be checked against a transcript of that
session to verify the guess. The situation is similar to many password-based
authentication and session-setup protocols. Encrypted key exchange protocols,
discovered by Bellovin and Merritt~\cite{BelM92}, do not suffer from this
problem. There a low entropy password is used to exchange a high entropy
secret that cannot efficiently be guessed using an off-line 
attack\footnote{Of course on-line attacks where all possible passwords are
  tried one by one can never be prevented.}.
Using encrypted key exchange protocols for basic access control
would strengthen the security of the passport considerably.

In Sect.~\ref{sec-newapps} we have seen that it will be inevitable
that other applications want to use the infrastructure available on the 
chip for other purposes than the original ones.
In the current system it is already possible to sign things with a private
key, but this causes some unwanted side effects as already described 
in Sect.~\ref{sec-newapps}.
In order to prevent this the standards should be rewritten in such a way that
at least these additional functions can be used and preferably in a disjoint
setting from the border inspection functions.
A possible implementation for this could be to have an X.509 certificate
included with a public key that has nothing to do with the MRZ or other
information needed for the border inspection tasks.

\bibliography{strings,passport}

\end{document}